\documentclass[10pt, final, journal, letterpaper, twocolumn]{IEEEtran}
\makeatletter
\def\ps@headings{%
\def\@oddhead{\mbox{}\scriptsize\rightmark \hfil \thepage}%
\def\@evenhead{\scriptsize\thepage \hfil \leftmark\mbox{}}%
\def\@oddfoot{}%
\def\@evenfoot{}}
\makeatother 
\IEEEoverridecommandlockouts

\usepackage{graphicx}
\usepackage{amsmath,amssymb}

\usepackage{algorithm}
\usepackage{algpseudocode}
\usepackage{amsmath}
\usepackage{graphics}
\usepackage{epsfig}

\usepackage[numbers,sort&compress]{natbib}
\usepackage{flafter}

\begin{document}
%
\title{Secrecy Rate Maximization with Outage Constraint in Multihop Relaying Networks}
\author{\IEEEauthorblockN{Jianping~Yao and
Yuan~Liu,~\IEEEmembership{Member,~IEEE}
}

\thanks
{
The authors are with the School of Electronic and Information Engineering, South China University of Technology, Guangzhou 510641, China (e-mails: yaojp\_scut@qq.com, eeyliu@scut.edu.cn).
}
}
\maketitle

\begin{abstract}
In this paper, we study the secure transmission in multihop wireless networks with randomize-and-forward (RaF) relaying, in the presence of randomly distributed eavesdroppers. By considering adaptive encoder with on-off transmission (OFT) scheme, we investigate the optimal design of the wiretap code and routing strategies to maximize the secrecy rate while satisfying the secrecy outage probability (SOP) constraint. We derive the exact expressions for the optimal rate parameters of the wiretap code. Then the secure routing problem is solved by revising the classical Bellman-Ford algorithm. Simulation results are conducted to verify our analysis.

\end{abstract}


\IEEEpeerreviewmaketitle

\section{Introduction}

\IEEEPARstart{N}{odes} cooperation to enhance the security communication for wireless network has attracted considerable interest. Most of the work in this area focused on one- or two-hop networks \cite{Dong2010,zheng2015outage,Zou2013,Cai2014,Mo2012,zhang2016energy,zhang2016artificial}. The authors in \cite{cai2013secure} studied the problem of secure connectivity for randomize-and-forward (RaF) relaying strategy in cooperative wireless networks.
The authors in \cite{el2016enhancing} and \cite{el2016physical} investigated the physical layer security of a buffer-aided half-duplex (HD) and full-duplex (FD) RaF relaying wireless network, respectively.
Only a few attempts have addressed this problem in multihop relaying networks. In multihop relaying networks, cooperation leads to smaller geographical distance in each hop, which decreases the path loss and destructive fading effects. On the other hand, more hops results in lower spectral efficiency, and more transmitting means more chances to be eavesdropped. As a result, there exists a tradeoff between the hops and secure performance, i.e., the problem of secure routing. The authors in \cite{sheikholeslami2016energy,ghaderi2015minimum} considered minimum energy routing in the presence of either multiple malicious jammers or eavesdroppers, to guarantee certain end-to-end performance. The authors in \cite{he2013end,lee2015full} considered the problem of how to communicate securely with the help of untrusted relays and full-duplex jamming relays, respectively. The authors in \cite{yjpjrnl2015} addressed the secure routing problem in multihop wireless networks with half-duplex decode-and-forward (DF) relaying, where the locations of the eavesdroppers were modeled as a homogeneous Poisson point process (PPP). The authors in \cite{yao2016secure} studied the secure connection problem in multihop wireless networks, where FD jamming relays operate to enhance the physical layer security.

To our best knowledge, the rate adaptation by wiretap code designs has not been considered in secure routing yet. In this paper, we study the secrecy rate maximization problem in multihop wireless networks with RaF relaying, in the presence of randomly distributed eavesdroppers. The location of the eavesdroppers is modeled as a homogeneous PPP. The locations and channel state information (CSI) of the eavesdroppers are unknown at the legitimate nodes. We consider the secrecy outage probability (SOP) of a selected path as a constraint for the system. We adopt the on-off transmission scheme \cite{zhou2013rethinking}, i.e., transmission occurs only when the received signal-to-noise (SNR) at the legitimate node exceeds the threshold. At each hop, a wiretap code is used at the legitimate node to ensure security according to the feedback of SNR from the receiver. For adaptive encoder with on-off transmission (OFT) scheme, we consider a network design problem of maximizing the secrecy rate by the joint design the rate parameters of the wiretap code and routing subject to the end-to-end SOP constraint. The explicit expressions for the optimal rate parameters for any given path are derived. Then the secure routing problem is solved by adopting the revised Bellman-Ford algorithm.

\section{System Model and Performance Metric}

We consider a multihop relaying network consisting of randomly distributed legitimate nodes and eavesdroppers. For a typical path ${L}$, the transmission of each hop is in a separate slot subject to multiple colluding eavesdroppers. The location of the eavesdroppers is modeled as a homogeneous PPP with density ${\lambda _e}$ denoted by ${\Phi _{ne}}\left( {n = 1, \ldots ,{\left| {L} \right|}}\right)$, where ${\left| {L} \right|}$ is the number of hops of path ${L}$. All nodes are equipped with one omni-directional antenna. The locations and CSI of the eavesdroppers are unknown at the legitimate nodes, but the legitimate nodes can characterize the statistics of the eavesdroppers' channel gains and positions. All channels are modeled by large-scale fading with path loss exponent $\alpha$ along with small-scale Rayleigh fading. The corresponding channel gains are independent exponentially distributed with unit mean. Hence, the instantaneous received SNR at the legitimate node and eavesdropper $e$ in ${\Phi _{ne}}\left( {n = 1, \ldots ,{\left| {L} \right|}} \right)$ of the $n$th-hop can be respectively given as:
\begin{align}
{{\tt SNR}_n} \triangleq \frac{{{p_n}{H_n}}}{{|{D_n}{|^\alpha }}},\;\;\;
{{\tt SNR}_{ne}} \triangleq \frac{{{p_n}{S_{ne}}}}{{|{X_{ne}}{|^\alpha }}},
\label{snr_hop}
\end{align}
where ${p_n}$ represents the transmit power of the $n$th-hop; ${H_{{n}}}$ and $|{D_{{n}}}|$ are Rayleigh fading gain and distance between the $n$th-hop, respectively; ${S_{{n}{e}}}$ and $|{X_{{n}{e}}}|$ are Rayleigh fading gain and distance between the legitimate node and eavesdropper $e$ of the $n$th-hop, respectively. Since RaF relaying strategy is used, the source and relays nodes use different codebooks to transmit the secret message and the relays add independent randomization in each hop when re-coding the received message, so the eavesdroppers can not combine the information from multiple hops. The total received SNR at the eavesdroppers of the $n$th-hop is given as
\begin{align}
{\tt SNR}_{ne}^{{\rm{sum}}} = \sum\limits_{e \in {\Phi _{ne}}} {{\tt SNR}_{ne}}
= \sum\limits_{e \in {\Phi _{ne}}} {\frac{{{p_n}{S_{ne}}}}{{|{X_{ne}}{|^\alpha }}}}.
\label{snr_RaF_1}
\end{align}


We adopt the well-known Wyner's encoding scheme as follows \cite{Wyner1975}. For each of the distinct messages to be secretly transmitted, the transmitters would randomly choose one of several possible codewords according to a local random number generator. We can therefore understand a wiretap code as having a nested structure. The transmitters use two kinds of parameters, namely, the rate of the transmitted codewords ${R_t}$ and the rate of the confidential information ${R_s}$ to encode the secrecy message. The rate difference ${R_e} \triangleq {R_t} - {R_s}$ represents the rate loss for transmitting the message securely against eavesdropping. We assume that there is no retransmission in each hop to reduce the risk of being eavesdropped. We adopt the OFT scheme, i.e., transmission occurs only when the received SNR at the legitimate node exceeds the SNR threshold ${\beta _t} = {2^{{R_s}}} - 1$. To enable the OFT scheme, the receiver needs to feed back full information of the instantaneous SNR to the transmitter. By doing so, the transmitter is able to adaptively choose ${R_t}$, i.e., the transmitter sets ${R_t}$ arbitrarily close to the capacity of the legitimate channel. Then the SOP of the $n$th-hop is defined as the conditional probability \cite{zhou2013rethinking,zheng2015outage}:
\begin{align}
{\mathcal{P}_{\textrm{so}}^{'}} \triangleq {\mathcal{P}}\left( {{{\log }_2}\left( \frac{1 + {\tt SNR}_n}{1 + {\tt SNR}_{ne}^{{\rm{sum}}}} \right) < {{R_s}}\mid {\tt SNR}_n>{\beta _t}} \right).
\label{P_so_1}
\end{align}

According to \cite{Koyluoglu2012}, the message is secured if every hop in the path is secured. Hence, the end-to-end SOP of the path can be expressed as
\small{\begin{align}
{\mathcal{P}_{\textrm{so}}} \triangleq \mathcal{P} \bigg( &\mathop {\min } \left\{{{\log}_2}\left(\frac {1 + {{\tt SNR}_n}}{1 + {\tt SNR}_{ne}^{{\rm{sum}}}} \right)\right\} < {R_s} \Big| \mathop {\min } \left\{ {{\tt SNR}_n} \right\}>{\beta _t} \bigg).
\label{P_so_path_1}
\end{align}
}
Since the transmission in each hop is independent, (\ref{P_so_path_1}) is equivalent to
\begin{align}
{\mathcal{P}_{\textrm{so}}} =1- \prod\limits_{n = 1}^{\left| {L} \right|}{\left(1-{\mathcal{P}_{\textrm{so}}^{'}}\right)}.
\label{P_so_path_2}
\end{align}

\newtheorem{theorem}{Theorem}
\begin{theorem}\label{Theorem_P_so}
The end-to-end SOP of a path is given by
\begin{align}
{\mathcal{P}_{{\rm{so}}}= 1 - \exp \left[ { - {K_1}{2^{\frac{{2{R_s}}}{\alpha }}}\sum\limits_{n = 1}^{\left| {L} \right|} {|{D_n}{|^2}} } \right]},
\label{P_so_RaF_final}
\end{align}
where ${K_1} = \pi {\lambda _e}\Gamma (1 + \frac{2}{\alpha })\Gamma (1 - \frac{2}{\alpha })$ and $\Gamma \left( \cdot \right)$ is the gamma function.

\begin{IEEEproof}
See Appendix \ref{appendices_Theorem_P_so}.
\end{IEEEproof}
\end{theorem}
\vspace{3ex}

According to (\ref{P_so_RaF_final}), we can see that ${\mathcal{P}_{{\rm{so}}}}$ is independent of the transmit power $p_n$, which means that the improvement of capacity for both the legitimate and eavesdropping channels are the same as the transmit power grows. Hence, we can not enhance the security by increasing the transmit power.

\section{Secrecy Rate Maximization}
In this study, we consider the optimization problem of maximizing the secrecy rate while satisfying the SOP constraint as follows:
\begin{subequations}
\begin{align}
\label{P1_a}\textbf{P1}:~~ \underset{R_s,{L \in {L}_{SD}}}{\max}\quad &\mathbb{C}_s = \frac{{R_s}}{\left| L \right|}, \\
\label{P1_b}s.t.\quad &{\mathcal{P}_{\textrm{so}}}\left(R_s,L\right)\leq \epsilon,\\
\label{P1_c}&R_s >0,
\end{align}
\end{subequations}
where ${L_{{SD}}}$ is the set of all possible paths $L$ connecting the source and destination; $\epsilon\in \left[0,1\right]$ represents the minimum security requirement.

From (\ref{P1_a}), we know that more hops ${\left| L \right|}$ reduces the transmission distance in each hop which in turn allows higher $R_s$ but the decrease of $\mathbb{C}_s$. In addition, more hops also means more chances to be eavesdropped which leads to higher ${\mathcal{P}_{\textrm{so}}}$. As a result, there exist a tradeoff between the hops ${\left| L \right|}$ and secure performance.

Since $R_s$ is independent of $L$ and ${\mathcal{P}_{\textrm{so}}}$ is an increasing function of $R_s$, it is easy to observe that the optimal solution occurs at ${\mathcal{P}_{\textrm{so}}}\left(R_s,L\right) = \epsilon$ for (\ref{P1_b}). The obtained $R_s$ is given by
\begin{align}
{{R_s}^* = \frac{\alpha }{2}{{\log }_2}\left( {\frac{{\ln \frac{1}{{1 - \epsilon }}}}{{{K_1}\sum\limits_{l \in L} {|{D_l}{|^2}} }}} \right)}.
\label{R_s_1}
\end{align}


Moreover, substituting (\ref{R_s_1}) into (\ref{P1_c}), we can obtain the upper bound of the eavesdroppers' density as
\begin{align}
\lambda_e < \frac{\ln \frac{1}{{1 - \epsilon }}}{\pi\Gamma (1 + \frac{2}{\alpha })\Gamma (1 - \frac{2}{\alpha }){{\sum\limits_{l \in {L}} {|{D_l}{|^2}}}}}.
\label{lambda_e_path}
\end{align}

Then we can reformulate $\textbf{P1}$ as
\begin{subequations}
\begin{align}
\label{P1.1_a}\textbf{P1}':~ \underset{{L \in {L}_{SD}}}{\max}\quad &\mathbb{C}_s = {\frac{\alpha }{{2\left| L \right|}}{{\log }_2}\left( {\frac{{\ln \frac{1}{{1 - \epsilon }}}}{{{K_1}\sum\limits_{l \in L} {|{D_l}{|^2}} }}} \right)}\\
\label{P1.1_b}s.t.\quad &\lambda_e < \frac{\ln \frac{1}{{1 - \epsilon }}}{\pi\Gamma (1 + \frac{2}{\alpha })\Gamma (1 - \frac{2}{\alpha }){{\sum\limits_{l \in {L}} {|{D_l}{|^2}}}}}.
\end{align}
\end{subequations}

\begin{theorem}\label{Theorem_Routing}
The optimization problem $\textbf{P1}'$ is equivalent to the following problem:
\begin{align}
\textbf{P2}:~ {{M}_t}({L^*}) = \mathop {\max}\limits_{1 \le v \le N_L - 1} {{M}_t}({\tilde L_v}),
\label{P2_s1}
\end{align}
where
\begin{align}
 {{M}_t}({\tilde L_v}) = {\frac{\alpha }{{2\left| {{{\tilde L}_v}} \right|}}{{\log }_2}\left( {\frac{{\ln \frac{1}{{1 - \epsilon }}}}{{{K_1}\sum\limits_{l \in {{\tilde L}_v}} {|{D_l}{|^2}} }}} \right)},
\label{P2_s2}
\end{align}

and

\begin{equation}
 {\tilde L_v} = \left\{
 \begin{array}{ll}
 \textbf{arg}~ \mathop {\min}\limits_{L \in {L_{SD}}:\left| L \right| \le v} {{\sum\limits_{l \in L} {|{D_l}{|^2}} }}, &\text{if}~{{\sum\limits_{l \in {{\tilde L}_v}} {|{D_l}{|^2}}}} < \frac{\ln \frac{1}{{1 - \epsilon }}}{K_1},\\
\varnothing, &\text{otherwise}.
\end{array}
\right.
\label{P2_s3}
\end{equation}

Here ${N_L}$ is the number of the legitimate nodes; ${ L^*}$ and ${\tilde L_v}$ are the optimal solution to $\textbf{P2}$ and subproblem (\ref{P2_s3}), respectively; ${{M}_t}({ L^*})$ and ${{M}_t}({\tilde L_v})$ are the corresponding optimal values of the objective function.

\begin{IEEEproof}
See Appendix \ref{appendices_Theorem_Routing}.
\end{IEEEproof}
\end{theorem}
\vspace{3ex}

 Theorem \ref{Theorem_Routing} implies that $\textbf{P1}'$ can be solved optimally by solving a sequence of subproblems (\ref{P2_s3}). It is easy to see that the solution to (\ref{P2_s3}) without constraint for a given hop-count $v$ means that each link uses ${|{D_l}{|^2}}$ as the link weights to find the path connecting the source and the destination which has the minimum total link weights and is no more than $v$ hops. Having the weighting factor of $v$ in the objective function, the optimization problem cannot be solved directly by using the classical Bellman-Ford algorithm, because it does not take the weighting factor into account. Hence, we develop a revised Bellman-Ford algorithm as shown in Algorithm \ref{alg:al1}.

\begin{algorithm}[!h]
\caption{Secure routing algorithm}
\label{alg:al1}
\begin{algorithmic}[1]
\Require
Topology information which contains the neighbor list and transmission distance ${|{D_l}{|}}$ between them;
\Ensure
\State Each legitimate node sets ${|{D_l}{|^2}}$ as link weights.
\State Obtain the shortest path ${\tilde L_v}$ in each iteration $v\left( {1, \ldots ,N_L-1} \right)$ while the total link weights is less than $\frac{\ln \frac{1}{{1 - \epsilon }}}{K_1}$  by the classical Bellman-Ford algorithm;
\State Calculate the function values for each path ${\tilde L_v}$ using (\ref{P2_s2});
\State Get the optimal path ${L^*}$ with the maximum function value using (\ref{P2_s1});
\State Get the optimal value of $R_s$ of the path ${L^*}$ using (\ref{R_s_1});\\
\Return ${L^*}$ and ${R_s}^*$;
\end{algorithmic}
\end{algorithm}

Before using the algorithm, each legitimate node calculates the distances between itself and all neighboring nodes. Then it sends its topology information to all neighboring nodes and runs this distributed algorithm separately \cite{bertsekas1992data}. From Algorithm \ref{alg:al1}, the main implementation complexity of the algorithm is dominated by Step 1. Hence, Algorithm \ref{alg:al1} has the same level of computational complexity as the classical Bellman-Ford algorithm, which is $O({N_L}^3)$ \cite{bertsekas1992data}.

\section{Numerical Results}

\begin{table*}[!t]
\centering
\caption{Comparison of Different Relaying Strategies Varying with the Number of Legitimate Nodes}
\label{tab:RaF_C_route}
\begin{tabular}{|c|c|c|c|c|c|c|c|c|c|c|}
\hline
$N_L$     &$10$  &$20$  &$30$  &$40$  &$50$  &$60$  &$70$  &$80$  &$90$  &$100$  \\ \hline
$\mathbb{\overline{C}}_s^{{\rm{RaF}}}$  &0.2382 &0.3373 &$0.3739$ &$0.3932$ &$0.4049$ &$0.4120$ &$0.4182$ &$0.4223$ &$0.4257$ &$0.4283$  \\ \hline
$\mathbb{\overline{C}}_s^{{\rm{DF}}}$  &$0$ &$0.00001$ &$0.0001$ &$0.008$ &$0.0031$ &$0.0072$ &$0.0127$ &$0.0194$ &$0.0256$ &$0.0311$  \\ \hline
\end{tabular}
\end{table*}

In this section, we illustrate and analyze the numerical results for the SOP performance and secrecy rate performance.

First, we simulate a multihop wireless network, in which the nodes are deployed in a $2000 \times 2000$ square area. The eavesdroppers are located at randomly and follow a homogeneous PPP. We consider an example of 6 legitimate nodes whose locations are at $\left(- 10,0\right)$, $\left(5\cos\left(0.75\pi\right),5\sin\left(0.75\pi\right)\right)$, $\left(0,0 \right)$, $\left(5\cos\left(- 0.25\pi\right),5\sin\left(- 0.25\pi\right)\right)$, $\left(10,0\right)$ and $\left(15\cos\left(0.25\pi\right),15\sin\left(0.25\pi\right) \right)$. We assume that $\alpha = 4$ and $p_n = 80 \textrm{dB}$ for all $n$.

\begin{figure}[!h]
\centering
  \includegraphics[width=6.5cm]{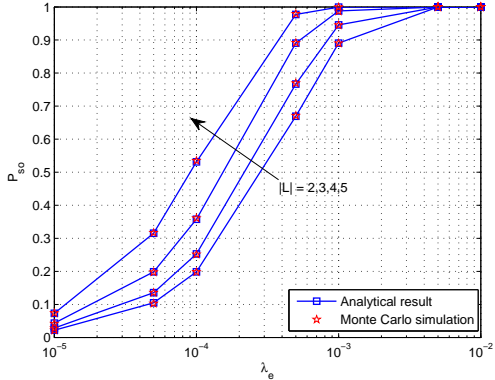}\\
  \caption{SOP versus the density of the eavesdroppers ${\lambda _e}$.}\label{fig:M_RF_C}
\end{figure}
Fig. \ref{fig:M_RF_C} depicts the SOP versus the density of the eavesdroppers ${\lambda _e}$. It can be seen that our analysis results perfectly agree to the Monte Carlo simulation results, which validates our analysis. As the value of ${\lambda _e}$ and the number of hops grows, the SOP increases.

\begin{figure}[!h]
\centering
  \includegraphics[width=6.5cm]{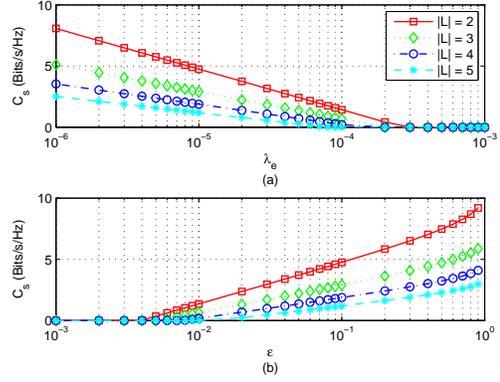}\\
  \caption{(a) Secrecy rate $\mathbb{C}_s$ versus the density of the eavesdroppers $\lambda_e$. (b) Secrecy rate $\mathbb{C}_s$ versus secrecy constraint $\epsilon$.}\label{fig:RaF_lambda_e_epsilon_routing}
\end{figure}
Fig. \ref{fig:RaF_lambda_e_epsilon_routing}(a) and Fig. \ref{fig:RaF_lambda_e_epsilon_routing}(b) depict the secrecy rate $\mathbb{C}_s$ versus the density of the eavesdroppers $\lambda_e$ and versus secrecy constraint $\epsilon$, respectively. As the density of the eavesdroppers $\lambda_e$ increases, SOP constraint $\epsilon$ decreases and as the number of hops grows, the cost of against eavesdropping increases. Hence, the secrecy rate $\mathbb{C}_s$ of the path and the corresponding tolerable density of the eavesdroppers decrease.

Then, we simulate a scenario that legitimate nodes are placed uniformly at random on a $50 \times 50$ square area in the center of the network. The source node is placed at the lower left corner of the network and the destination is located at the upper right corner. Note that the eavesdroppers are still randomly distributed in the entire network of size $2000 \times 2000$. Our goal is to find the highest secrecy rate path between the source and destination. We assume that $\alpha = 4$, $\lambda_e = 10^{-5}$, $\epsilon = 0.1$ and $p_n = 80 \textrm{dB}$ for all $n$. For different number of legitimate nodes, we simulate the different routing algorithms $10000$ times and obtain the average secrecy rates. The results are shown in Table \ref{tab:RaF_C_route}.

In Table \ref{tab:RaF_C_route}, $N_L$ denotes the number of legitimate nodes. $\mathbb{\overline{C}}_s^{{\rm{RaF}}}$ and $\mathbb{\overline{C}}_s^{{\rm{DF}}}$ represent the average secrecy rate for RaF and DF relaying strategies, respectively\footnote{Because of space limitation, the details of the analysis for the DF scheme are not presented in the paper.}. As shown in the table, both of them increase with the number of legitimate nodes growing. It is because that more legitimate nodes results in the more choices to get a safer route for a given source-destination pair. And the secrecy performance of RaF scheme is always better than that of the DF scheme.

\section{Conclusion}

In this paper, we investigated the secure transmission in a multihop relaying network with the help of the relays using randomize-and-forward (RaF) relaying strategy in the presence of homogeneous Poisson point process (PPP) distributed colluding eavesdroppers. We formulated the problem of maximizing the secrecy rate by the joint design of the wiretap code and routing  subject to the secrecy outage probability (SOP) constraint. Explicit expressions for the optimal rate parameter of wiretap code were derived, and the secure routing problem was also solved by the revised Bellman-Ford algorithm.

\appendices
\section{Proof of Theorem \ref{Theorem_P_so}}\label{appendices_Theorem_P_so}

According to (\ref{snr_hop}) and (\ref{snr_RaF_1}), (\ref{P_so_1}) can be rewritten as
\begin{align}
{\mathcal{P}_{\textrm{so}}^{'}} = 1- {\mathcal{P}}\left( {{{\log }_2}\left( \frac{1 + \frac{{{p_n}{H_n}}}{{|{D_n}{|^\alpha }}}} {1 + \sum\limits_{e \in {\Phi _{ne}}} {\frac{{{p_n}{S_{ne}}}}{{|{X_{ne}}{|^\alpha }}}} } \right) > {R_s}\mid {\tt SNR}_n>{\beta _t}} \right).
\label{P_so_2_2}
\end{align}

Then, (\ref{P_so_2_2}) can be turned to
\begin{align}
{\mathcal{P}_{\textrm{so}}^{'}} = 1- {\mathcal{P}}\left( {{\frac{{{p_n}{H_n}}}{{|{D_n}{|^\alpha }}} >{\beta _t}+ {{2^{{R_s}}}\sum\limits_{e \in {\Phi _{ne}}} {\frac{{{p_n}{S_{ne}}}}{{|{X_{ne}}{|^\alpha }}}} }}\mid {\tt SNR}_n>{\beta _t}} \right).
\label{P_so_2}
\end{align}

Since the memoryless property of exponential distribution, (\ref{P_so_2}) is equivalent to
\begin{align}
{\mathcal{P}_{\textrm{so}}^{'}} = 1- {\mathcal{P}}\left( {{\frac{{{H_n}}}{{|{D_n}{|^\alpha }}} >{{2^{{R_s}}}\sum\limits_{e \in {\Phi _{ne}}} {\frac{{{S_{ne}}}}{{|{X_{ne}}{|^\alpha }}}} }}} \right).
\label{P_so_3}
\end{align}

Then (\ref{P_so_3}) can be rewritten as
\begin{align}
{\mathcal{P}_{\textrm{so}}^{'}} = 1- {\mathbb{E}_{\mathop {{\Phi _{ne}},{S_{ne}}} }}\left\{ \exp \left[-{ {{|{D_n}{|^\alpha }}{{2^{{R_s}}}\sum\limits_{e \in {\Phi _{ne}}} {\frac{{{S_{ne}}}}{{|{X_{ne}}{|^\alpha }}}} }}} \right] \right\}.
\label{P_so_4}
\end{align}

Since ${S_{ne}}$ is independent and identically distributed, thus the expectation over the sum of ${S_{ne}}$  is equal to the product of the expectation over ${S_{ne}}$. Then (\ref{P_so_4}) can be turned to
\begin{align}
{\mathcal{P}_{\textrm{so}}^{'}} = 1- {\mathbb{E}_{\mathop {{\Phi _{ne}},{S_{ne}}} }}\left\{\prod\limits_{e \in {\Phi _{ne}}} \exp \left[-{ {{|{D_n}{|^\alpha }}{{2^{{R_s}}} {\frac{{{S_{ne}}}}{{|{X_{ne}}{|^\alpha }}}} }}} \right] \right\}.
\label{P_so_5}
\end{align}

Then (\ref{P_so_5}) can be rewritten to
\begin{align}
{\mathcal{P}_{\textrm{so}}^{'}} = 1-{\mathbb{E}_{\mathop {{\Phi _{ne}}} }\left\{ {{\prod\limits_{e \in {\Phi _{ne}}} {\frac{1}{{\frac{{{2^{{R_s}}}|{D_n}{|^\alpha }}}{{|{X_{ne}}{|^\alpha }}} + 1}}} } } \right\}}.
\label{P_so_6}
\end{align}

For a homogeneous PPP, the probability generating functional (PGFL) is given by \cite{Chiu2013}
\begin{align}
{\mathbb{E}_{{\Phi _e}}}\left[ {\prod\limits_{{{e}} \in {\Phi _e}} {f\left( {{e}} \right)} } \right] = \exp \left[ { - {\lambda _e}\int_{{\mathbb{R}^2}} {1 - f\left( {{e}} \right)  {\rm{d}}{{e}}} } \right].
\label{PGFL}
\end{align}

According to (\ref{PGFL}), (\ref{P_so_6}) can be rewritten as:
\begin{align}
{\mathcal{P}_{\textrm{so}}^{'}} = 1-{ {\exp \left[ { - {\lambda _e}\int_{{\mathbb{R}^2}} {\frac{{{2^{{R_s}}}|{D_n}{|^\alpha }}}{{{2^{{R_s}}}|{D_n}{|^\alpha } + |{X_{ne}}{|^\alpha }}}} \, \text{d}e} \right]} }.
\label{P_so_7}
\end{align}

Then (\ref{P_so_7}) can be simplified as:
\begin{align}
{\mathcal{P}_{\textrm{so}}^{'}} = 1- {\exp \left[ { - {K_1}{2^{\frac{{2{R_s}}}{\alpha }}}{|{D_n}{|^2}} } \right]}.
\label{P_so_final}
\end{align}

Replacing ${\mathcal{P}_{\textrm{so}}^{'}}$ with (\ref{P_so_final}) into (\ref{P_so_path_2}), we can obtain (\ref{P_so_RaF_final}).

\section{Proof of Theorem \ref{Theorem_Routing}}\label{appendices_Theorem_Routing}
Similar to \cite{yjpjrnl2015}, we can prove that problem $\textbf{P1}'$ without the constraint (\ref{P1.1_b}) can be solved optimality by solving a sequence of subproblems
\begin{align}
 {{M}_t}({ L_v}) = {\frac{\alpha }{{2\left| {{{ L}_v}} \right|}}{{\log }_2}\left( {\frac{{\ln \frac{1}{{1 - \epsilon }}}}{{{K_1}\sum\limits_{l \in {{ L}_v}} {|{D_l}{|^2}} }}} \right)},~ v= {1, \ldots ,N_L-1},
\label{P2_tr1}
\end{align}
where ${ L_v}$ is the optimal solution to following problem
\begin{align}
 {L_v} =
 \textbf{arg}~ \mathop {\max}\limits_{L \in {L_{SD}}:\left| L \right| \le v} {\frac{\alpha }{2v}{{\log }_2}\left( {\frac{{\ln \frac{1}{{1 - \epsilon }}}}{{{K_1}\sum\limits_{l \in {{ L}_v}} {|{D_l}{|^2}} }}} \right)}.
\label{P2_tr2}
\end{align}

 Then, (\ref{P2_tr2}) is equivalent to
\begin{align}
 {L_v} =
 \textbf{arg}~ \mathop {\min}\limits_{L \in {L_{SD}}:\left| L \right| \le v} {{\sum\limits_{l \in L} {|{D_l}{|^2}} }}.
\label{P2_tr3}
\end{align}

When ${L_v}$ is not satisfied the constraint (\ref{P1.1_b}), the problem (\ref{P2_tr3}) has no feasible solution for the problem $\textbf{P1}'$.

This completes the proof.

\footnotesize
\bibliographystyle{IEEEtran}
\bibliography{myreference}
\end{document}